\documentclass[review]{elsarticle}

\usepackage{lineno,hyperref}
\usepackage{amssymb,amsmath}
\usepackage{graphicx}
\usepackage{float}
\usepackage{subfig}
\usepackage{multirow}
\usepackage{array}
\usepackage{lscape}
\usepackage{subfloat}
\usepackage{rotating}
\usepackage{caption}
\usepackage{color}
\usepackage{epstopdf}
\usepackage{threeparttable}
\usepackage{verbatim}
\usepackage{chngpage}
\usepackage{amsmath}
\usepackage{algorithm}
\usepackage[noend]{algpseudocode}
\usepackage{subfig}
\usepackage[table, dvipsnames]{xcolor}
\usepackage{multirow, booktabs}
\usepackage{threeparttable}
\usepackage{booktabs,makecell}

\makeatletter
\def\BState{\State\hskip-\ALG@thistlm}
\makeatother
\allowdisplaybreaks

\modulolinenumbers[5]

\journal{}

\usepackage[numbers]{natbib}

\begin{document}

\begin{frontmatter}

\title{An intelligent financial portfolio trading strategy using deep Q-learning}

\author{Hyungjun Park}
\address{Department of Industrial and Management Engineering, Pohang University of Science and Technology, 77 Cheongam-Ro, Nam-Gu, Pohang, Gyeongbuk 37673, Rep. of Korea}

\author{Min Kyu Sim}
\address{Department of Industrial and Systems Engineering, Seoul National University of Science and Technology, 232, Gongneung-Ro, Nowon-Gu, Seoul, 01811, Rep. of Korea}

\author{Dong Gu Choi\corref{mycorrespondingauthor}}
\cortext[mycorrespondingauthor]{Corresponding author}
\address{Department of Industrial and Management Engineering, Pohang University of Science and Technology, 77 Cheongam-Ro, Nam-Gu, Pohang, Gyeongbuk 37673, Rep. of Korea}
\ead{dgchoi@postech.ac.kr}

\begin{abstract}
Portfolio traders strive to identify dynamic portfolio allocation schemes so that their total budgets are efficiently allocated through the investment horizon. This study proposes a novel portfolio trading strategy in which an intelligent agent is trained to identify an optimal trading action by using deep Q-learning. We formulate a Markov decision process model for the portfolio trading process, and the model adopts a discrete combinatorial action space, determining the trading direction at prespecified trading size for each asset, to ensure practical applicability. Our novel portfolio trading strategy takes advantage of three features to outperform in real-world trading. First, a mapping function is devised to handle and transform an initially found but infeasible action into a feasible action closest to the originally proposed ideal action. Second, by overcoming the dimensionality problem, this study establishes models of agent and Q-network for deriving a multi-asset trading strategy in the predefined action space. Last, this study introduces a technique that has the advantage of deriving a well-fitted multi-asset trading strategy by designing an agent to simulate all feasible actions in each state. To validate our approach, we conduct backtests for two representative portfolios and demonstrate superior results over the benchmark strategies.
\end{abstract}


\end{frontmatter}

\section{Introduction}

A goal of financial portfolio trading is maximizing the trader's monetary wealth by allocating capital to a basket of assets in a portfolio over the periods during the investment horizon. Thus, portfolio trading is the most important investment practice in the buy-side financial industry. Portfolio traders strive to establish trading strategies that can properly allocate capital to financial assets in response to time-varying market conditions. Typical objective functions for trading strategy optimization include expected returns and the Sharpe ratio (i.e., risk-adjusted returns). In addition to optimizing an objective function, a trading strategy should achieve a reasonable turnover rate so that it is applicable to real-world financial trading. If the turnover rate is not reasonable, transaction costs hurt overall trading performance.

Portfolio trading is an optimization problem that involves a sequential decision-making process across multiple rebalancing periods. In this process, the stochastic components of time-varying market variables should be considered. Thus, the problem of deriving an optimal portfolio trading strategy has traditionally been formulated as a stochastic optimization problem~\citep{Consigli:1998,Golub:1995,Kouwenberg:2001}. To handle these stochastic components over multiple periods, most related studies have developed heuristic methods~\citep{Brock:1992,Chen:2017,Chen:2016,Derigs:2003,Leigh:2002,Papailias:2015,Zhang:2015,Zhu:2009}. In very recent years, reinforcement learning (RL) has become another popular approach for financial portfolio trading problems. In RL methods, a learning agent can understand a complex financial environment by attempting various trading actions and revising its trading action policy, and then optimize their trading strategies based on these experiences. In addition, these methods have the important advantage that learning agents can update their trading strategies based on their experiences on future trading days. Instead of simply maintaining trading strategies derived from historical data, learning agents can adapt their strategies using their observed experiences on each real trading day~\citep{Wang:2016}. With these advantages and the increasing popularity of RL algorithms, many previous studies have been conducted to apply RL algorithms to various portfolio trading problem settings~\citep{Almahdi:2017,Almahdi:2019,Bertoluzzo:2012,Casqueiro:2006,Dempster:2006,Deng:2016,Eilers:2014,Jeong:2019,Jiang:2017,Moody:2001,Moody:1998,Neuneier:1996,Neuneier:1998,O:2006,Pendharkar:2018,Zhang:2015}.

According to the recent evolution of RL methods, some researchers~\citep{Deng:2016,Jeong:2019,Jiang:2017} have started to use deep RL (DRL) methods, which is the combination of RL and deep neural network (DNN), to overcome the unstable performance of previous RL methods. Nevertheless, we believe that this line of study is yet to mature in terms of practical applicability because of the following two reasons. First, most studies based on DRL methods focus on a single-asset trading~\citep{Deng:2016,Jeong:2019}. Because most traders generally have multiple securities, additional decision-making steps are necessary even though single asset trading rules are derived. Second, even if a study dealing with the multi-asset portfolio trading, the actions determine portfolio weights~\citep{Jiang:2017}. The action spaces that determine the portfolio weights cannot be followed directly because it requires additional decisions on how to satisfy the target weight. We will explain this issue in more detail in Section~\ref{subsec:Literature Review3}.

To overcome the current limitations, this study proposes a new approach for deriving a multi-asset portfolio trading strategy using deep Q-learning, one of the most popular DRL methods. In this study, we focus on a multi-asset trading strategy and define an intuitive trading action set that can be interpreted as direct investment guides to traders. In the action space used in this study, each action includes trading directions corresponding to each asset in a portfolio, and each trading direction comprises either holding each asset or buying or selling each asset at a prespecified trading size. Although a recent study~\citep{Pendharkar:2018} argues that optimizing a trading strategy based on a discrete action space has a negative effect, we find that our discrete action space modeling allows for a lower turnover rate and is more practical than continuous action space modeling is.

To develop a practical multi-asset trading strategy, this study tackles a few challenging aspects.
First, setting a discrete combinatorial action space may lead to infeasible actions, and, thus, we may derive an unreasonable trading strategy (i.e., a strategy with frequent and pointless portfolio weight changes that only leads to more transaction costs). To address this issue, we introduce a mapping function that enables the agent to prevent the selection of unreasonable actions by mapping infeasible actions onto similar and valuable actions. By applying this mapping function, we can derive a reasonable trading strategy in the practical action space.
Second, the action space that determines a trading direction for each asset in the portfolio has a dimensionality problem~\citep{Moody:2001}. As the number of assets in portfolio increasing, then the size of action space increases exponentially because a trading agent must determine a combination of trading directions for several assets in the portfolio. This is why previous studies related to this action space have only considered a single asset or a single risky asset with a risk-free asset trading. Therefore, in this study, we overcome this limitation and conduct the first study of multi-asset trading in the practical action space by using DQL.
Third, although we use years of financial data, these data may not provide enough training data for the DRL agent to learn a multi-asset trading strategy in the financial environment. Because learning a strategy mapping from a joint state to a joint action is necessary lots of data. There is a given amount of data, so we need to make the agent gains more experience within the training data and learns as much as possible. Thus, we achieve sufficient learning by simulating all feasible actions in each state and then updating the agent's trading strategy using the learning experiences from the simulation results. This technique allows the agent to gain and learn enough experience to derive a well-fitted multi-asset trading strategy.

The rest of this paper is organized as follows. In Section 2, we first review the related literature and present the differences between our study and previous studies. Section 3 describes the definition of our problem, and Section 4 introduces our approach for deriving an intelligent trading strategy. In Section 5, we provide experimental results to validate the advantages of our approach. Finally, we conclude in Section 6 by providing relevant implications and identifying directions for future research. 

\section{Literature Review}

Portfolio trading is an optimization problem that involves a sequential decision-making process over multiple rebalancing periods. In addition, the stochastic components of market variables should be considered in this process. Thus, traditionally, the derivation of portfolio trading strategies has been formulated as a stochastic programming problem to find an optimal trading strategy. Recently, much effort has been made to solve this stochastic optimization problem using a learning-based approach, RL. To formulate this stochastic optimization problem, it is necessary to determine how to measure the features of the stochastic components corresponding to changes in the financial market. Utilizing technical indicators is more common than utilizing the fundamental indexes of securities in daily frequency portfolio trading, as in our study.

This section reviews how previous studies have attempted to model stochastic market components to formulate the portfolio trading problem and derive an optimal trading strategy. Section~\ref{subsec:Literature Review1} provides a brief description of previous studies that formulate the stochastic components of the financial market. Section~\ref{subsec:Literature Review2} reviews previous studies that discuss heuristic methods for deriving an optimal trading strategy. Section~\ref{subsec:Literature Review3} reviews previous studies that address the stochastic optimization problem to derive an optimal trading strategy using RL.

\subsection{Stochastic programming-based models}\label{subsec:Literature Review1}
Early studies on portfolio trading and, sometimes, management used stochastic programming-based models. Stochastic programming models formulate a sequence of investment decisions over time that can maximize a portfolio manager's expected utility up to the end of the investment horizon. \citet{Golub:1995} modeled an interest rate series as a binomial lattice scenario using Monte Carlo procedures to solve a money management problem with stochastic programming. \citet{Kouwenberg:2001} solved an asset-liability management problem using the event tree method to generate random stochastic programming coefficients. \citet{Consigli:1998} used scenario-based stochastic dynamic programming to solve an asset-liability management problem. However, stochastic programming-based models have the limitation of needing to generate numerous scenarios to solve a complex problem, such as understanding a financial environment, resulting in a large computational burden.

\subsection{Heuristic methods}\label{subsec:Literature Review2}
Because of this limitation of stochastic programming-based models, many studies have devised heuristic methods (i.e., trading heuristics). One of the most famous such methods is technical analysis for asset trading. This method provides a simple and sophisticated way to identify hidden relationships between market features and asset returns through the study of historical data. Using these identified relationships, investments are made in assets by taking appropriate positions. \citet{Brock:1992} conducted backtests with real and artificial data using moving average and trading range strategies. \citet{Zhu:2009} considered theoretical rationales for using technical analysis and suggested a practical moving average strategy to determine a portion of investments. \citet{Chourmouziadis:2016} suggested an intelligent stock-trading fuzzy system based on rarely used technical indicators for short-term portfolio trading. 
Another popular heuristic method is the pattern matching (i.e., charting heuristics) method, which detects critical market situations by comparing the current series of market features to meaningful patterns in the past. \citet{Leigh:2002} developed a trading strategy using two types of bull flag pattern matching. \citet{Chen:2016} proposed an intelligent pattern-matching model based on two novel methods in the pattern identification process. 
The other well-known heuristic method is a metaheuristic algorithm that can find a near optimal solution in acceptable computation time. \citet{Derigs:2003} developed a decision support system generator for portfolio management using simulated annealing, and \citet{Potvin:2004} applied genetic programming to generate trading rules automatically. \citet{Chen:2017} used a genetic algorithm to group stocks with similar price series to support investors in making more efficient investment decisions. 
However, these heuristic methods have limited ability to fully search a very large feasible solution space because they are inflexible. Thus, we need to be careful about the reliability of obtaining an optimal trading strategy using these methods.

\subsection{Reinforcement learning-based methods}\label{subsec:Literature Review3}
A recent research direction is optimizing a trading strategy using RL such that a learning agent develops a policy while interacting with the financial environment. Using RL, a learning-based method, the learning agent can search for an optimal trading strategy flexibly in a high-dimensional environment. Unlike supervised learning, RL allows learning from experience, leading to training the agent with unlabeled data obtained from interactions with the environment. 

In the earliest such studies, \citet{Neuneier:1996,Neuneier:1998} optimized multi-asset portfolio trading using Q-learning, a model-free and value-based RL. In other early studies, \citet{Moody:1998} and \citet{Moody:2001} used \emph{Direct} RL with \emph{Recurrent} RL as a base algorithm and derived a multi-asset long-short portfolio trading strategy and a single-asset trading rule, respectively. \emph{Direct} RL is policy-based RL, which optimizes an objective function by adjusting policy parameters, and \emph{Recurrent} RL is an RL algorithm in which the last action is received as an input. These studies introduced several measures, such as the Sharpe ratio and the Sterling ratio, as objective functions and compared the trading strategies derived using different objectives. \citet{Casqueiro:2006} derived a single-asset trading strategy using Q-learning, which can maximize the Sharpe ratio. \citet{Dempster:2006} developed an automated foreign exchange trading system using an adaptive learning system with a base algorithm of \emph{Recurrent} RL by dynamically adjusting a hyper-parameter depending on the market situation. \citet{O:2006} proposed a Q-learning-based local trading system that categorized an asset price series into four patterns and applied different trading rules. \citet{Bertoluzzo:2012} suggested a single-asset trading system using Q-learning with linear and kernel function approximations. \citet{Eilers:2014} developed a trading rule for an asset with a seasonal price trend using Q-learning. \citet{Zhang:2015} derived a trading rule generator using extended classifier systems combined with RL and a genetic algorithm. \citet{Almahdi:2017} suggested a \emph{Recurrent} RL-based trading decision system that enabled multi-asset portfolio trading and compared the performance of the system when several different objective functions were adopted. \citet{Pendharkar:2018} suggested an indices trading rule derived using two different RL methods, on-policy (SARSA) and off-policy (Q-learning) methods and compared the performance of these two methods, and it also compared the performances of discrete and continuous agent action space modeling. \citet{Almahdi:2019} used a hybrid method that combined \emph{Recurrent} RL and particle swarm optimization to derive a portfolio trading strategy that considers real-world constraints.

More recently, DRL, which combines deep learning and RL algorithms, was developed, and, thus, studies have suggested using DRL-based methods to derive portfolio trading strategies. DRL methods enable an agent to understand a complex financial environment through deep learning and to learn a trading strategy by automatically applying an RL algorithm. \citet{Jiang:2017} used a deep deterministic policy gradient (DDPG), an advanced method of combining policy-based and value-based RL, and introduced various DNN structures and techniques to trade a portfolio consisting of cash and several cryptocurrencies. \citet{Deng:2016} derived an asset trading strategy using a \emph{Recurrent} RL-based algorithm and introduced a fuzzy deep recurrent neural network that used fuzzy representation to reduce uncertainty in noisy asset prices and used a deep recurrent neural network to consider the previous action and utilize high-dimensional nonlinear features. \citet{Jeong:2019} derived an asset trading rule that determined actions for assets and the number of shares for the actions taken. To learn this trading rule, \citet{Jeong:2019} used a deep Q-network (DQN) with a novel DNN structure consisting of two branches, one of which learned action values while the other learned the number of shares to take to maximize the objective function.

The above studies used various RL-based methods in different problem settings. All of the methods performed well in each setting, but some issues limit the applicability of these methods to the real world. First, some problem settings did not consider transaction costs~\citep{Bertoluzzo:2012,Eilers:2014,Jeong:2019,O:2006,Pendharkar:2018}. A trading strategy developed without assuming transaction costs is likely to be impractical for application to the real world. The second issue is that some strategies consider trading for only one asset~\citep{Almahdi:2017,Bertoluzzo:2012,Casqueiro:2006,Dempster:2006,Eilers:2014,Deng:2016,Jeong:2019,Moody:2001,Zhang:2015}. A trading strategy of investing in only one risky asset may have high risk exposure because it has no risk diversification effect. Finally, in previous studies deriving multi-asset portfolio trading strategies using RL, the agent's action space was defined as the portfolio weights in the next period~\citep{Almahdi:2017,Almahdi:2019,Jiang:2017,Moody:1998}. The action spaces of these studies do not provide portfolio traders with a direct guide that is applicable to a real-world trading scenario that includes transaction costs. This is because there are many different ways to transition from the current portfolio weight to the next portfolio weight. Thus, previous studies using portfolio weights as the action space required finding a way to minimize transaction costs at each rebalancing moment. Rebalancing in a way that reduces both transaction costs and dispersion from the next target portfolio is not an easily solved problem~\citep{Grinold:2000}. In addition, a portfolio trading strategy derived based on the action spaces of the previous studies may be difficult to apply to real-world trading because the turnover rate is likely to be high. An action space that determines portfolio weights can result in frequent asset switching because the amount of asset changes has no upper bound. Thus, we contribute to the literature by deriving a portfolio trading strategy that has no such issues.

\section{Problem definition}\label{sec:Problem}

In this study, we consider a portfolio consisting of cash and several risky assets. All assets in the portfolio are bought using cash, and the value gained from selling assets is held in cash. That is, the agent cannot buy an asset without holding cash and cannot sell an asset without holding the asset. This type of portfolio is called a long-only portfolio, which does not allow short selling. Our problem setting also has a multiplicative profit structure in that the portfolio value accumulates based on the profits and losses in previous periods. We consider proportional transaction costs that are charged according to a fixed proportion of the amount traded in transactions involving buying or selling. In addition, we allow the agent to partially buy or sell assets (e.g., the agent can buy or sell half of a share of an asset).

We set up some assumptions in our problem setting. First, transactions can only be carried out once a day, and all transactions in a day are made at the closing price in the market at the end of that day. Second, the liquidity of the market is high enough that each transaction can be carried out immediately for all assets. Third, the trading volume of the agent is very small compared to the size of the whole market, so the agent's trades do not affect the state transition of the market environment.

To apply RL to solve our problem, we need a model of the financial environment that reflects the financial market mechanism. Using the notations summarized in Table~\ref{tab:Notation}, we formulate a Markov decision process (MDP) model that maximizes the portfolio return rate in each period by selecting sequential trading actions for the individual assets in the portfolio according to time-varying market features (Table~\ref{tab:Feature}). 

\begin{table}[p]
\centering
\footnotesize
\caption{Summary of notations}\label{tab:Notation}
\begin{tabular}{l m{10cm}}
    \hline
    \multicolumn{2}{l}{\textbf{Decision variables}}\\
    $a_t=(a_{t,1},a_{t,2},...,a_{t,I})$ & agent's action at the end of period $t$ $\{a_t{\in}\mathbb{Z}^I:a_{t,i}{\in}\{-1,0,1\}\hspace{0.1cm}\forall i\}$\\\\
    \multicolumn{2}{l}{\textbf{Set and indices}} \\
    $i=0,1,2,...,I$ & portfolio asset index (i=0 represents cash) \\
    $t$ & time period index\\
    $S^{-}(a_t)$ & set of an index of selling assets when an agent takes action $a_t$ \par (i.e., $\{i{\in}\mathbb{Z}\hspace{0.1cm}|\hspace{0.1cm}0<i\leq{I}, a_{t,i}=-1\}$)\\
    $S^{+}(a_t)$ & set of an index of buying assets when an agent takes action $a_t$ \par (i.e., $\{i{\in}\mathbb{Z}\hspace{0.1cm}|\hspace{0.1cm}0<i\leq{I}, a_{t,i}=1\}$)\\\\
    \multicolumn{2}{l}{\textbf{Parameters}}\\
    $n$ & size of the time window containing recent previous market features \\
    $P_t$ & portfolio value changed by the action at the end of period $t$ \\
    $P_t'$ & portfolio value before the agent takes an action at the end of period $t$ \\
    $P_t^{s}$ & portfolio value at the end of period $t$ when the agent takes no action at the end of the previous period $t-1$ (static portfolio value in period $t$)\\
    $w_{t,i}$ & proportion of asset $i$ changed by the action at the end of period $t$\\
    $w_{t,i}'$ & proportion of asset $i$ before the agent takes an action at the end of period $t$\\
    $\hat{w}_{t,i}'$ & auxiliary parameter used to derive $w_{t,i}$\\
    $c_t$ & decay rate of transaction costs at the end of period $t$ \\
    $c^{-}$ & transaction cost rate for selling \\
    $c^{+}$ & transaction cost for buying \\
    $\delta$ & trading size for selling or buying $\biggl(0<\delta<\frac{P_t'}{I}\biggr)$\\
    $\rho_t$ & return rate of the portfolio in period $t$ $\biggl(=\frac{P_{t}-P_{t-1}}{P_{t-1}}\biggr)$\\
    $o_{t,i}$ & opening price of asset $i$ in period $t$\\
    $p_{t,i}$ & closing price of asset $i$ in period $t$\\
    $h_{t,i}$ & highest price of asset $i$ in period $t$\\
    $l_{t,i}$ & lowest price of asset $i$ in period $t$\\
    $v_{t,i}$ & volume of asset $i$ in period $t$\\
    \hline
\end{tabular}
\end{table}
\begin{table}
\centering
\footnotesize
\caption{Summary of market features}\label{tab:Feature}
\begin{tabular}{l m{10cm}}
    \hline
        \multicolumn{2}{l}{\textbf{Features}}\\
    $k_{t,i}^{c}$ & rate of change of the closing price of asset $i$ in period $t$
    $\biggl(=\frac{p_{t,i}-p_{t-1,i}}{p_{t-1,i}}\biggr)$ \\
    $k_{t,i}^{o}$ & ratio of the opening price in period $t$ to the closing price in period $t-1$ for asset $i$ $\biggl(=\frac{o_{t,i}-p_{t-1,i}}{p_{t-1,i}}\biggr)$ \\
    $k_{t,i}^{h}$ & ratio of the closing price to the highest price of asset $i$ in period $t$ $\biggl(=\frac{p_{t,i}-h_{t,i}}{h_{t,i}}\biggr)$ \\
    $k_{t,i}^{l}$ & ratio of the closing price to the lowest price of asset $i$ in period $t$ $\biggl(=\frac{p_{t,i}-l_{t,i}}{l_{t,i}}\biggr)$ \\
    $k_{t,i}^{v}$ & rate of change of the volume of asset $i$ in period $t$ $\biggl(=\frac{v_{t,i}-v_{t-1,i}}{v_{t-1,i}}\biggr)$ \\
    \hline
\end{tabular}
\end{table}

\subsection{State space}\label{subsec:state space}

The state space of the agent is defined as the weight vector of the current portfolio before the agent selects an action and the tensor that contains the market features (technical indicators) for the assets in the portfolio. This type of state space is similar to that used in a previous study~\citep{Jiang:2017}. That is, the state in period $t$ can be represented as below (Equations (1)-(3)):
\begin{align}\label{eq:statevariables}
& s_{t} = (X_t,w_t') \text{, } \\
& w_t' = (w_{t,0}',w_{t,1}',w_{t,2}',...,w_{t,I}')^{T} \text{, } \\
& X_{t} = [K_{t}^{c},K_{t}^{o},K_{t}^{h},K_{t}^{l},K_{t}^{v}] \text{, }
\end{align}
where $w_t'$ denotes the weight vector of the current portfolio and $X_t$ represents the technical indicator tensor for the assets in the portfolio. For this tensor, we use five technical indicators for the assets in the portfolio, as below (Equations (4)):
\begin{align}\label{eq:technicalindicatorvector}
& k_{t}^{x} = (k_{t,1}^{x},k_{t,2}^{x},...,k_{t,I}^{x})^{T} \hspace{1cm} \forall x \in \{c,o,h,l,v\} \text{, }
\end{align}

\begin{figure*}
	\centering
	\includegraphics[width=0.3\textwidth]{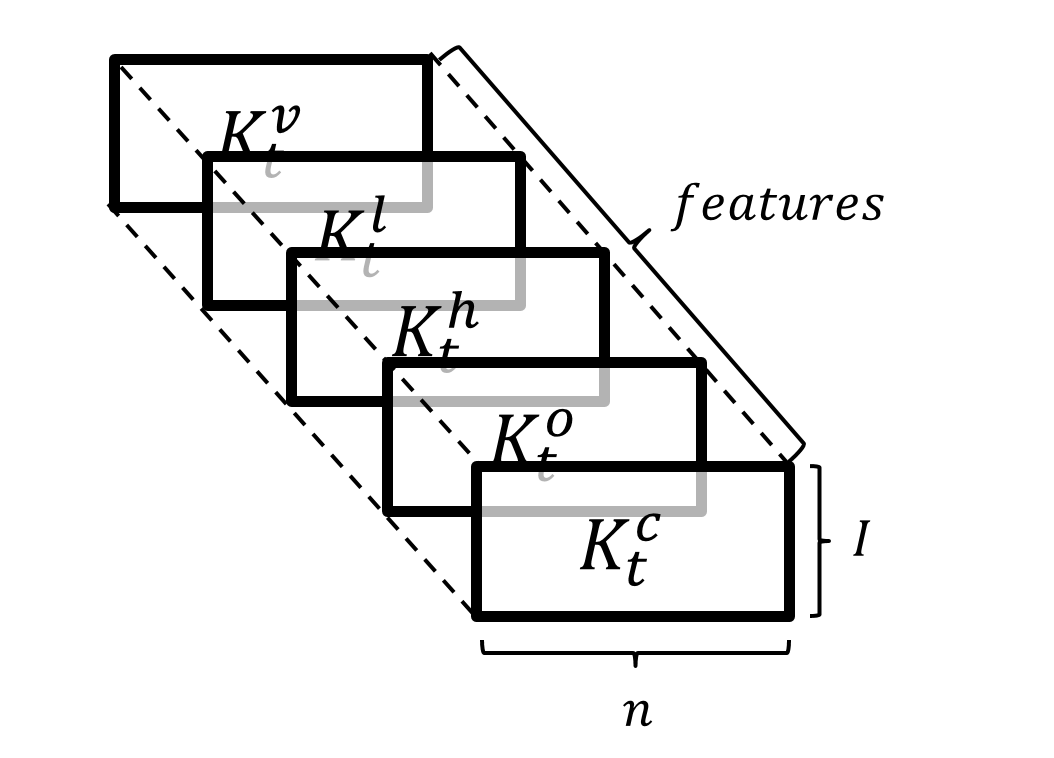}
	\caption{Market feature tensor ($X_t$)}\label{fig:state}
	\end{figure*}

Every set of five technical indicators can be expressed as a matrix (Equations (5)), where the rows represent each asset in the portfolio and the columns represent the series of recent technical indicators in the time window. Here, if we set a time window of size $n$ (considering $n$-lag autocorrelation) and a portfolio of $I$ assets, the technical indicator tensor is an $(I,n,5)$-dimensional tensor, as in Figure \ref{fig:state}. 

\begin{align}\label{eq:technicalindicatormaxtirx}
& K_{t}^{x}=[k_{t-n+1}^{x}|k_{t-n+2}^{x}|...|k_{t}^{x}] \hspace{1cm} \forall x \in \{c,o,h,l,v\}\text{, }
\end{align}

\subsection{Action space}\label{subsec:action space}

We define the action space to overcome the limitations of the action spaces in previous studies. Agent actions determine which assets to hold and which assets to sell or buy at prespecified a constant trading size. For example, if a portfolio includes two assets and the trading size is 10,000 $USD$, then the agent can select the action of buying 10,000 $USD$ of \textit{asset1} and selling 10,000 $USD$ of \textit{asset2}. The action space includes the trading directions of buying, selling, or holding each asset in the portfolio, so the action space contains $3^{I}$ different actions. These actions are expressed in a vector form that includes trading directions for each asset in a portfolio. In addition, each trading direction $(sell,hold,buy)$ is encoded as $(-1,0,1)$, respectively. For example, an action that involves selling \textit{asset1} and buying \textit{asset2} can be encoded into the vector $(-1,1)$.

Because the trading actions for individual assets are carried out in fixed trading size, this action space is modeled as a discrete type. Although this discrete action space may not be able to derive a trading strategy that outperforms trading strategies derived using a continuous action space~\citep{Pendharkar:2018}, this action space can provide a direct trading guide that a portfolio trader can follow in the real world. Furthermore, this discrete action space can derive a portfolio trading strategy with lower turnover relative to the strategies developed in previous studies. In previous studies, if a portfolio with a very large amount of capital is changed by a small amount in portfolio weight then the trader may pay significant transaction costs. In addition, the losses from these transaction costs can be very high because portfolio weight changes have no upper bound. In contrast, our action space has an upper bound for portfolio weight changes, and, thus, the issue of massive changes in portfolio weights and the resulting large losses from transaction costs do not arise. Our agent action space has these advantages, and the only disadvantage of the fixed trading amount is similar to the restrictions of hedge funds that allow portfolio traders to trade below a certain amount each day. Thus, our discrete agent action space is not too unrealistic to apply to real-world trading.   

\subsection{MDP modeling}\label{subsec:mdp}

With the state space and action space defined in the previous subsections, we can define the MDP model as follows. The financial market environment operates according to this model during the investment horizon. To define the transitions in the financial market environment (i.e., the system dynamics in the MDP model), we need to define following parameters and equations: 

\begin{align}\label{eq:weighttransition1}
& w_t = (w_{t,0},w_{t,1},w_{t,2},...,w_{t,I})^{T} \text{, } \\
& w_t\cdot\Vec{1}=w_t'\cdot\Vec{1}=1 \hspace{1cm} \forall t  \text{, } \\
& P_{t}'=P_{t-1}{w_{t-1}\cdot{\phi(k_{t}^{c})}} \hspace{1cm} \forall t  \text{, } \\
& w_t' = \frac{w_{t-1}\odot{\phi(k_{t}^{c})}}{w_{t-1}\cdot{\phi(k_{t}^{c})}} \hspace{1cm} \forall t \text{, }
\end{align}
where $w_t$ denotes the portfolio weight after the agent takes an action at the end of period $t$ (Equation (6)). Equation (7) provides the constraint that the portfolio weight elements sum to one in all periods. Equations (8) and (9) represent the change in the portfolio value and the change in the proportions of the assets in the portfolio given the changes in the value of each asset in the portfolio, respectively. Here, $\odot$ represents the elementwise product of two vectors, and $\Vec{1}$ is a vector of size \textit{I+1} with all elements equal to one. $\phi(\cdot)$ is an operator that not only increases a vector's dimension by positioning zero as the first element but also adds it to the $\Vec{1}$ vector ($\phi$ : $(e_1,e_2,...,e_I)^{T} \rightarrow (1,{e_1}+1,{e_2}+1,...,{e_I}+1)^{T}$).

\begin{figure*}[!htb]
	\centering
	\includegraphics[width=1.0\textwidth]{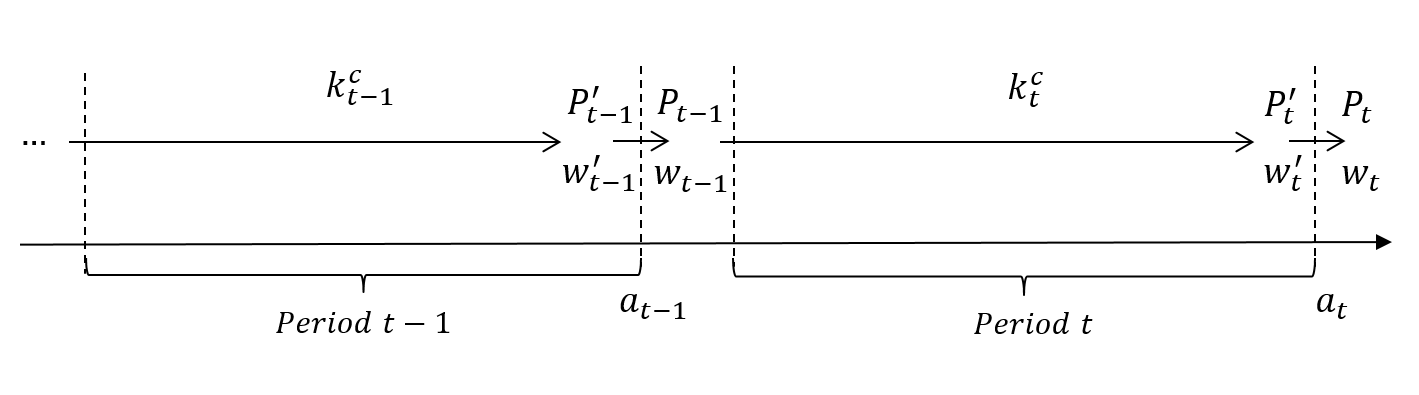}
	\caption{Financial environment transition}\label{fig:environment}
	\end{figure*}

Now, we can define the state changes after the agent takes an action as follows: 
\begin{align}\label{eq:weighttransition3}
& c_t = \frac{\delta}{P_t'}\biggl(c^{-}{\mid}S^{-}(a_t){\mid} +c^{+}{\mid}S^{+}(a_t){\mid}\biggr) \hspace{1cm} \forall t \text{, }\\
& P_t = P_t'(1-c_t) \hspace{1cm} \forall t \text{, } \\
& \hat{w}_t' = (\hat{w}_{t,0}',\hat{w}_{t,1}',\hat{w}_{t,2}',...,\hat{w}_{t,I}')^{T} \text{, } \\
& \hat{w}_{t,i}'=\begin{cases} w_{t,i}'-\frac{\delta}{P_t'} \hspace{0.2cm}if \hspace{0.2cm}i \in S^{-}(a_t),\\ w_{t,i}'+\frac{\delta}{P_t'} \hspace{0.2cm}if \hspace{0.2cm} i\in S^{+}(a_t),\\ w_{t,i}' \hspace{1.0cm}otherwise \end{cases} \forall_{i = 1 \dots I} \text{, } \\
& \hat{w}_{t,0}'=w_{t,0}'+\frac{\delta}{P_t'}\biggl((1-c^{-}){\mid}S^{-}(a_t){\mid}-(1+c^{+}){\mid}S^{+}(a_t){\mid}\biggl) \text{, } \\
& w_{t}=\frac{\hat{w}_{t}'}{\hat{w}_{t}'\cdot \overrightarrow{1}} \text{,}
\end{align}
After the agent takes an action, transaction costs arise, and the portfolio value is then decayed (Equations (10)-(11)). Here, ${\mid}S{\mid}$ is the size of set $S$. $\hat{w}_{t,i}'$ denotes the auxiliary weight of the portfolio that is needed to connect the change in the portfolio weights before and after the agent takes an action at the end of period $t$ (Equation (12)). The procedure by which the action selected by the agent is handled for trading in the financial environment is as follows. The auxiliary weight of an asset in the portfolio increases (or decreases) as a proportion of the trading size when buying (or selling) the asset. On the contrary, the auxiliary weights of the assets do not change when the agent holds the assets (Equation (13)).
As a result of selling asset, the proportion of cash increases by the proportion of the trading size discounted by the selling transaction cost rate. As a result of buying asset, the proportion of cash decreases by the proportion of the trading size multiplied by the buying transaction cost rate (Equation (14)). To ensure that the sum of the portfolio weight elements equals one after the agent takes an action, a process for adjusting the auxiliary weights is required (Equation (15)). In summary, the financial market environment transition is illustrated by Figure \ref{fig:environment}.

Last, the reward in the MDP model should reflect the contribution of the agent's action to the portfolio return. This reward can be simply defined as the portfolio return. However, if the portfolio return is defined only as a reward, then different reward criteria can be given depending on the market trend. For example, when the market trend is sufficiently improving, then no matter how poor the agent's action is, a positive reward is provided to the agent. In contrast, if the market trend is sufficiently negative, then no matter how helpful the agent's action is, a negative reward is provided to the agent. Thus, the reward must be defined as the rate of change in the portfolio value by which the market trend is removed. Therefore, we define the reward as the change in the portfolio value at the end of the next period relative to the static portfolio value (Equation (16)). The static portfolio value is the next portfolio value when the agent takes no action at the end of the current period (Equation (17)).
\begin{align}\label{eq:reward}
& r_{t}=\frac{P_{t+1}'-P_{t+1}^{s}}{P_{t+1}^{s}} \text{, }\\
& P_{t+1}^{s}=P_{t}'{w_{t}'\cdot{\phi(k_{t+1}^{c})}} \text{, }
\end{align}

\section{Methodology}\label{sec:Methodology}

In this section, we introduce our proposed approach for deriving the portfolio trading strategy using DQL. In our action space, some issues may prohibit a DQL agent from deriving an intelligent trading strategy. We first explain how to resolve these issues by introducing some techniques and applying existing methodologies. Then, we describe our DQL algorithm with these techniques. 


\subsection{Mapping function}\label{subsec:mapping}

In our action space, some actions are infeasible in some states (e.g., the agent cannot buy assets because of a cash shortage or cannot sell assets because of a shortage of held assets). To handle infeasible actions, we first set the action values (i.e., Q-values) of infeasible actions to be very low to mask these actions~\citep{Lanctot:2017}. Thus, we need to define a rule for selecting the appropriate action from the remaining actions when infeasible actions are excluded. For this rule, a simple way in which the agent selects the largest Q-valued one among remaining actions can be considered~\citep{Xiong:2018}. However, this simple rule can result in an unreasonable trading strategy. For example, when an agent's strategy selects the action of selling both \textit{asset1} and \textit{asset2} but this action is infeasible owing to a lack of \textit{asset2}, the action of buying both \textit{asset1} and \textit{asset2}, which is the largest Q-value action in the remaining action space, is selected. Because learning the similarity between actions is difficult for an RL agent, the agent will take this action without any doubt even though this selected action is the opposite of the original action determined by the agent's strategy. This issue leads to the selection of unreasonable actions, which degrades the trading performance. A mapping rule is required to map infeasible actions to similar and valuable actions in the feasible action set. Thus, we resolve this issue by introducing a mapping function that contains several mapping rules.

The mapping function is a type of constraint on action space in RL that allows the agent to derive a reasonable trading strategy. \citet{Pham:2018,Bhatia:2018} handled constrained action space by adding an optimization layer, so-called \emph{OptLayer}, for solving quadratic programming at the last layer of the agent's policy network, determining an action that minimizes differences from the output at the previous layer while satisfying constraints. However, the method cannot be applied directly to our situation because they can only deal with continuous action space. Although \emph{OptLayer} can be applied for handling our constrained action space by revising the integer quadratic programming (IQP) version, the solution of IQP suffers multiple-choice issue in this situation because there are tied actions which have the smallest distance from the original action. To overcome the limitation of the \emph{OptLayer}, we devise the mapping function based on a heuristic searching method. The function maps to one feasible action which has the largest Q-value among tied actions that have the smallest distance from the original action. Moreover, the computation costs of the mapping function are lower than those of \emph{OptLayer}. Therefore, the mapping function is an extended efficient method from the line of study to handle infeasible action using the concept of distance between actions, such as \emph{OptLayer}.

\begin{figure}[!htb]%
    \centering
    \subfloat[]{{\includegraphics[width=1.0\textwidth]{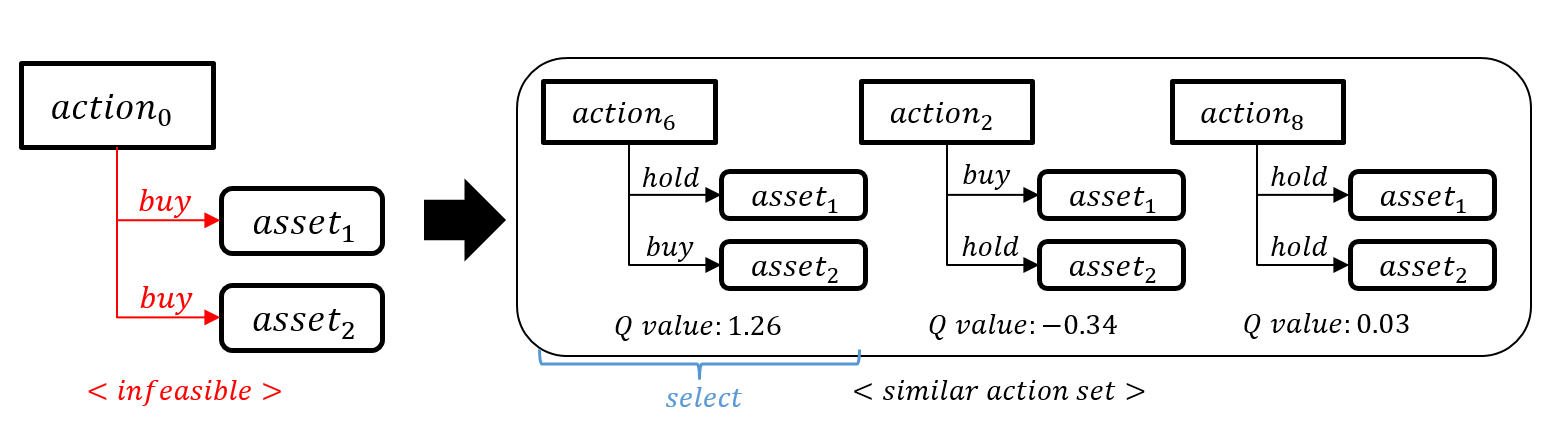}}}%
    \qquad
    \subfloat[]{{\includegraphics[width=1.0\textwidth]{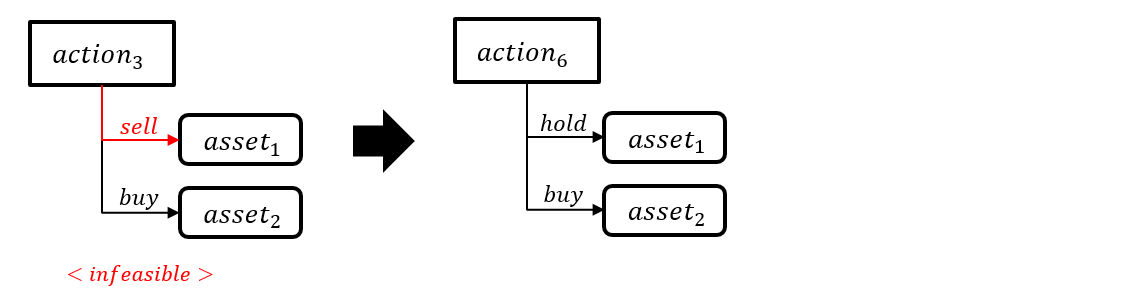}}}%
    \caption{Mapping examples of (a) a cash shortage and (b) an asset shortage}\label{fig:mapping}%
\end{figure}

The mapping function contains two mapping rules, each of which is required for mapping infeasible actions, that are divided into two cases. In the first case, the amount of cash is not sufficient to take an action that involves buying assets. In this case, a similar action set is derived by holding rather than buying a subset of the asset group to be bought in the original action. Thereafter, infeasible actions are mapped to the most valuable feasible actions in the similar action set. For example, if the action of buying both \textit{asset1} and \textit{asset2} is infeasible owing to a cash shortage, this action is mapped to the most valuable feasible action within the set of similar actions, which includes the action of buying \textit{asset1} and holding \textit{asset2}, the action of holding \textit{asset1} and buying \textit{asset2}, and the action of holding both \textit{asset1} and \textit{asset2}. In the second case, an action that involves selling assets is infeasible because of a shortage of the assets. In this case, the original action is simply mapped to an action in which the assets that are not enough to sell are held. These examples are illustrated in Figure \ref{fig:mapping}.

\begin{figure*}[!htb]
	\centering
	\includegraphics[width=1.0\textwidth]{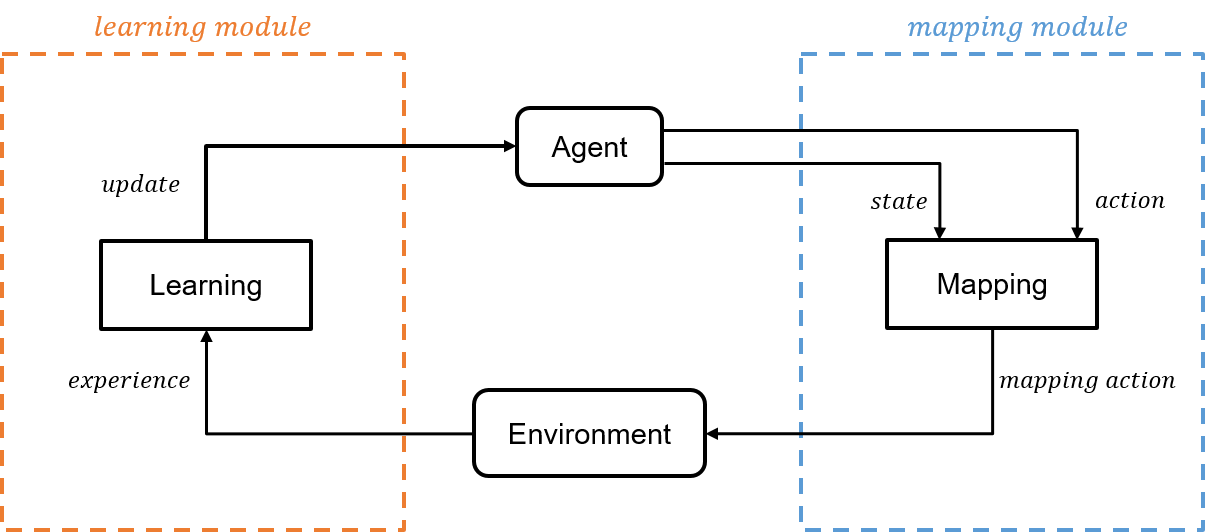}
	\caption{RL flow chart with mapping function}\label{fig:flow}
	\end{figure*}

We provide the details of the two mapping rules and the mapping function in the following pseudocode in Algorithm (1). In the Algorithm, the last part (i.e., Lines (26)-(27)) of the mapping rule for the second case(\textproc{Rule2}) is necessary. Because, in the second case, converting the original action of selling assets that cannot be sold into an action that holds the selling assets which cannot be sold, then the cash amount gained from selling assets is removed, causing the first infeasible action case to arise. Furthermore, this part of the code can handle the special case in which an asset shortage and a cash shortage occur simultaneously. Next, the RL flow chart with the mapping function technique is shown in Figure \ref{fig:flow}.

\begin{algorithm}
\caption{Mapping function}
\label{alg:mapping rule}
\begin{algorithmic}[1]
\State $s_t$: state of the agent
\State $a_t$: infeasible action in state $s_t$
\State $Q(s_t,a_t)$: Q-value for state action pair $(s_t,a_t)$
    \Procedure{Map}{$s_t$, $a_t$}
        \If{asset shortage for action $a_t$ in state $s_t$} 
            \State $a_{map} \gets \textproc{Rule2}(s_t,a_t)$
        \ElsIf{cash shortage for action $a_t$ in state $s_t$}
            \State $a_{map} \gets \textproc{Rule1}(s_t,a_t)$
        \EndIf
        \State \textbf{return} $a_{map}$
    \EndProcedure
    \Procedure{Rule1}{$s_t$, $a_t$}
    \State $MAXQ \gets -inf$
    \State subset\ of\ buying\ asset\ index:\ $S=\{C_1,...\}$
    \For{each subset $C$ in $S$}
        \State replicate\ action\ $\hat{a_t} \gets a_t$
        \For{each asset $j$ in $C$}
            \State $\hat{a}_{t,j} \gets 0$
            \If{converted action $\hat{a}_t$ is feasible in state $s_t$} 
                \If{$Q(s_t,\hat{a}_t) > MAXQ$}
                    \State $MAXQ \gets Q(s_t,\hat{a}_t)$
                    \State $a_{best} \gets \hat{a}_t$
                \EndIf
            \EndIf
        \EndFor
    \EndFor
    \State \textbf{return} $a_{best}$
    \EndProcedure
    \Procedure{Rule2}{$s_t$, $a_t$}
    \For{asset $i$ = 1,2...}
        \If{action $a_t$ to asset $i$ in state $s_t$ infeasible} 
            \State $a_{t,i} \gets 0$
        \EndIf
    \EndFor
    \If{converted action $a_t$ is infeasible in state $s_t$}
        \State $a_t \gets \textproc{Rule1}(s_t,a_t)$
    \EndIf
    \State \textbf{return} $a_t$
    \EndProcedure
\end{algorithmic}
\end{algorithm}

\subsection{DQN algorithm}\label{subsec:DQN}

We optimize the multi-asset portfolio trading strategy by applying the DQN algorithm. DQN is the primary algorithm for DQL. \citet{Mnih:2013} developed the DQN algorithm, and \citet{Mnih:2015} later introduced additional techniques and completed this algorithm. The base algorithm for DQN, Q-learning, is value-based RL, which is a method that approximates an action value (i.e., a Q-value) in each state. Further, Q-learning is a model-free method such that even if the agent does not have knowledge of the environment, the agent can develop a policy using repeated experience by exploring. In addition, Q-learning is an off-policy algorithm, that is, the action policy for selecting the agent's action is not the same as the update policy for selecting an action on the target value. An algorithm based on Q-learning that approximates the Q-function using DNN is the basis of DQN~\citep{Mnih:2013}. To prevent DNN from learning only through the experience of a specific situation, experience replay was introduced to sample a general experience batch from memory. Additionally, the DQN algorithm used two separate networks: a Q-network that approximates the Q-function and a target network that approximates the target value needed for the Q-network updated to follow a fixed target~\citep{Mnih:2015}. Based on this algorithm, we introduce several techniques to support the derivation of an intelligent trading strategy.

The existing DQN algorithm updates the Q-network with experience by allowing the agent to take only one action in each stage. Because the agent has no information about the environment, only one action is taken then proceeding to the next state. Thus, it is impossible to take multiple actions in the existing DQN. However, for this problem, we use historical technical indicator data of the assets in the portfolio as training data. Thus, our agent can take multiple actions in one state in each stage and observe all of their experiences based on those actions. To utilize this advantage, we introduce a technique that simulates all feasible actions in one state at each stage and updates the trading strategy by using the resulting experiences from conducting these simulations. 

\begin{figure}[!htb]%
    \centering
    \subfloat[]{{\includegraphics[width=0.7\textwidth]{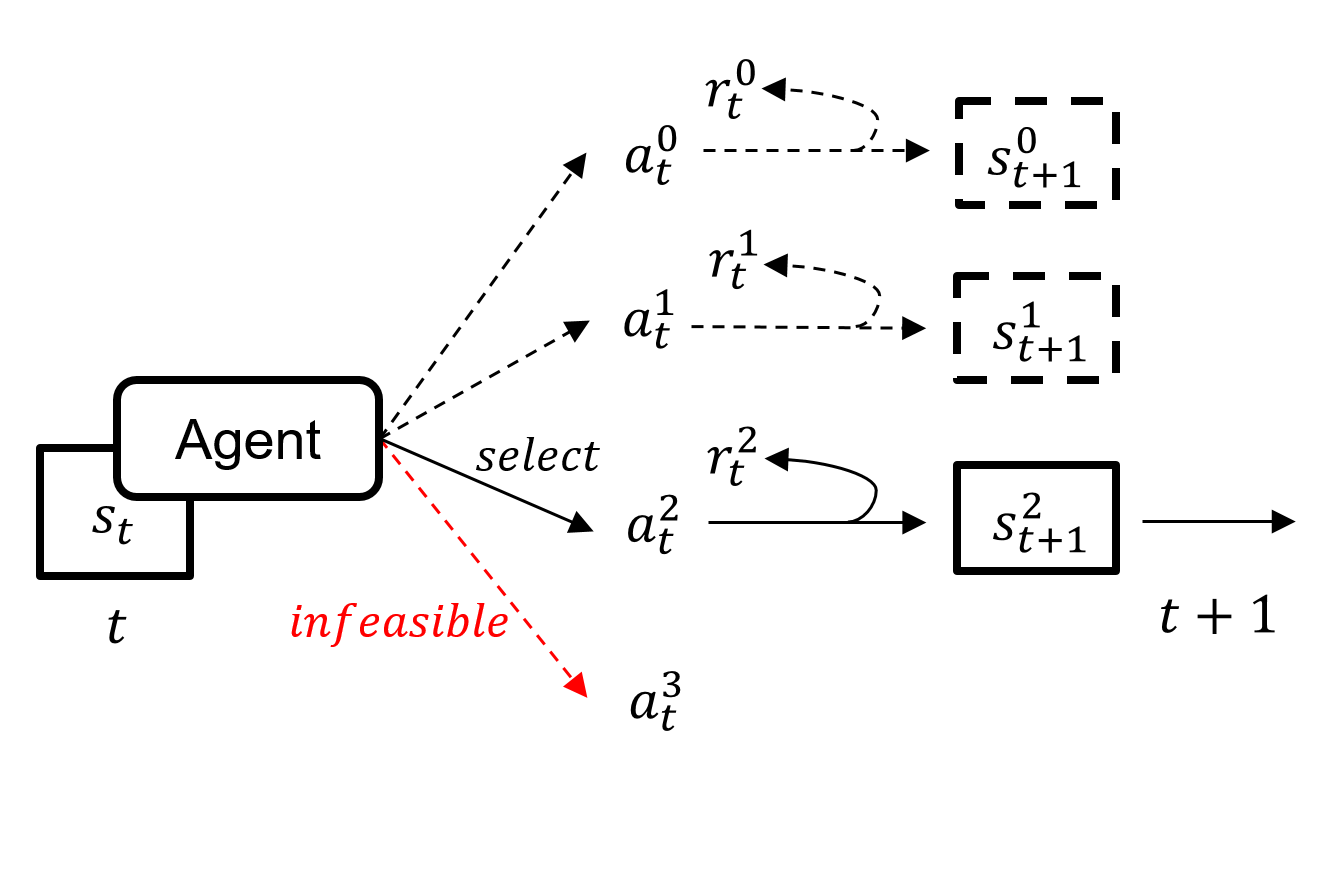}}}%
    \qquad
    \subfloat[]{{\includegraphics[width=0.5\textwidth]{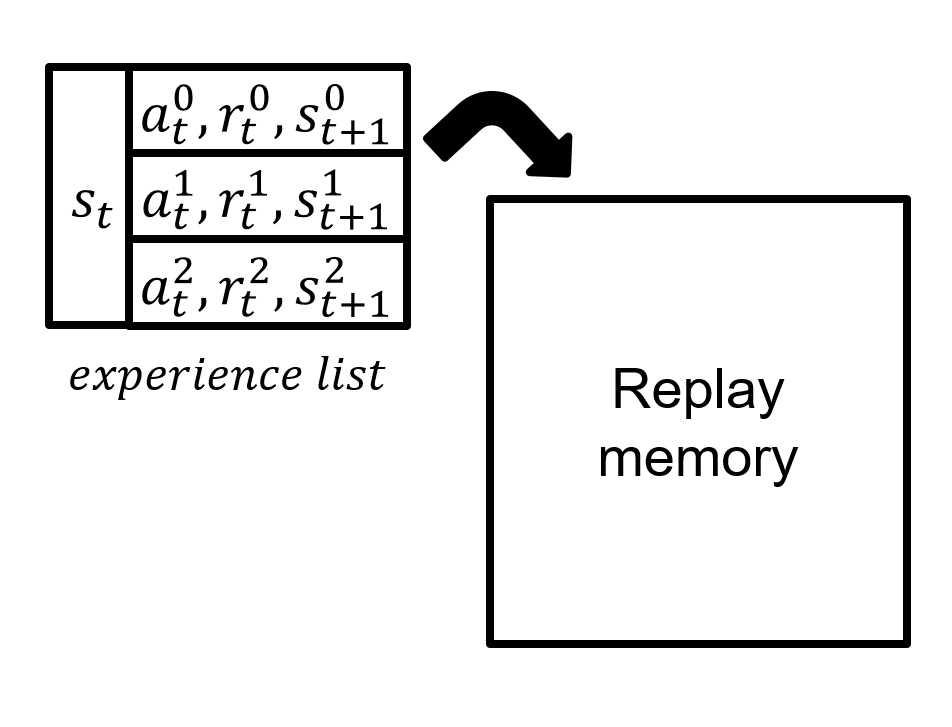}}}%
    \caption{(a) Simulating feasible actions, (b) Data structure for experience list}\label{fig:simul}%
\end{figure}

Motivated by \citet{Tan:2009}, we utilize a simulation technique that takes all feasible actions virtually to force to the agent learns about many experiences efficiently for deriving a fully searched multi-asset trading strategy. Thus, this technique can relax the data shortage issue that arises when deriving a multi-asset trading strategy. Although simulating all feasible actions can result in a huge computational burden, using multi-core parallel computing can prevent this computational burden from greatly increasing. Moreover, even if the agent takes multiple actions in the current state, the next state only depends on the action selected by the action policy (epsilon-greedy) with the mapping rule. The application of this technique requires a change in the data structure of the element in replay memory for storing a list of experiences in a state. The concepts related to this technique are illustrated in Figure \ref{fig:simul}. In this figure, $a_{t}^j$ means that the $j-th$ action of the agent is taken at the end of period $t$. $r_{t}^j$ is the reward obtained by taking action $a_{t}^j$, and $s_{t+1}^j$ is the next state that results from taking action $a_{t}^j$.

\begin{figure*}[!htb]
	\centering
	\includegraphics[width=1.0\textwidth]{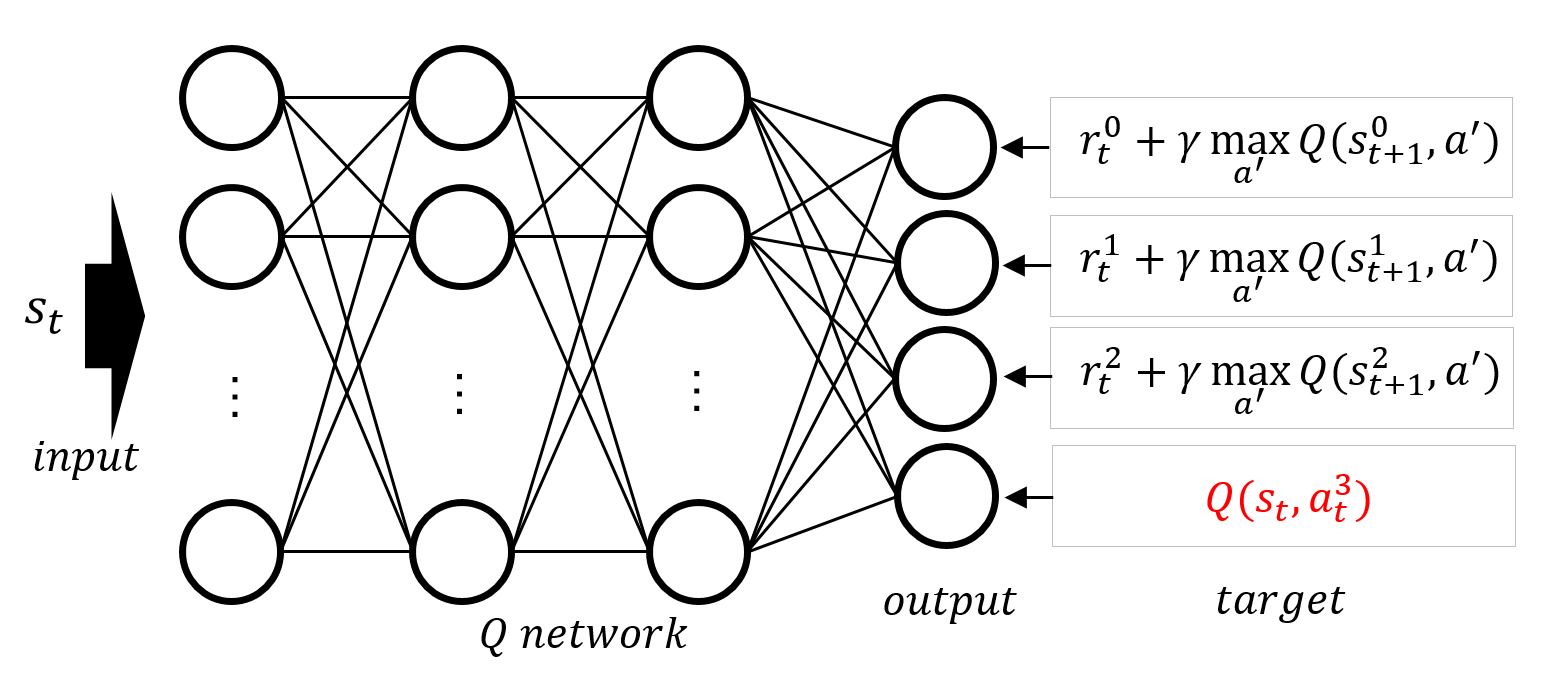}
	\caption{Updating a multiple output Q-network using an experience list}\label{fig:update}
	\end{figure*}

In DQN, a multiple output neural network is commonly adopted as the Q-network structure. In this network structure, the input of the neural network is the state, and the output is the Q-value of each action. Using the above technique, we can approximate the Q-value of all feasible actions by updating this multiple output Q-network in parallel with the experience list. To maintain Q-values of infeasible actions, the current Q-value of an infeasible state-action pair is assigned to the target value of the Q-network output of the corresponding infeasible action to set a temporal difference error of zero. Furthermore, as in DQN, several experience lists are sampled from replay memory, and the Q-network is updated using the experience list batch. A detailed description of the process for updating the Q-network is shown in Figure \ref{fig:update}.

In addition, to apply RL, learning episodes must be defined for the agent to explore and experience the environment. Rather than defining all of the training data, which cover several years, as one episode, we divide the training data into several episodes. If we define a much longer training episode than the investment horizon of the test data that will be used to test the trading strategy, this difference in the lengths of the training and test data can produce negative results. For example, in our experiment, the training and testing processes begin with the same portfolio weights. In this case, the farther the agent is from the beginning of the long training episode, the farther the agent is from the initial portfolio weights. Thus, it is difficult for the agent to utilize the critical experience obtained from the latter half of the long episode in the early testing process. Therefore, we divide training data into sets of the same length as the investment horizon of the test data (i.e., one year, as the investment horizon of the test data is a year in our experiment). Thus, the criteria for dividing the training data are defined in yearly units so that the episodes do not overlap (e.g., \textit{episode1} contains data from 2016, and \textit{episode2} contains data from 2015). In each training epoch, the agent explores and learns in an episode sampled from the training data. 

It is well known that more recent historical data have more explainable for predicting future data than less recent historical data have. Thus, it is reasonable to assign higher sampling probabilities to episodes that are closer to the test data period~\citep{Jiang:2017}. We use a truncated geometric distribution to assign higher sampling probabilities to episodes that are closer to the test period. This truncated geometric sampling distribution is expressed in Equation (18). Here, $y$ is the year of the episode, $y_v$ is the year of the test data, and $N$ is the number of total training episodes. $\beta$ is a parameter for this sampling distribution that ranges from zero to one. If this parameter is closer to one, episodes closer to the test period are sampled frequently.

\begin{align}\label{eq:model8}
& g_{\beta}(y)=\frac{\beta(1-\beta)^{y_v-y-1}}{1-(1-\beta)^{N}} \text{, }
\end{align}

\begin{figure*}[!htb]
	\centering
	\includegraphics[width=1.0\textwidth]{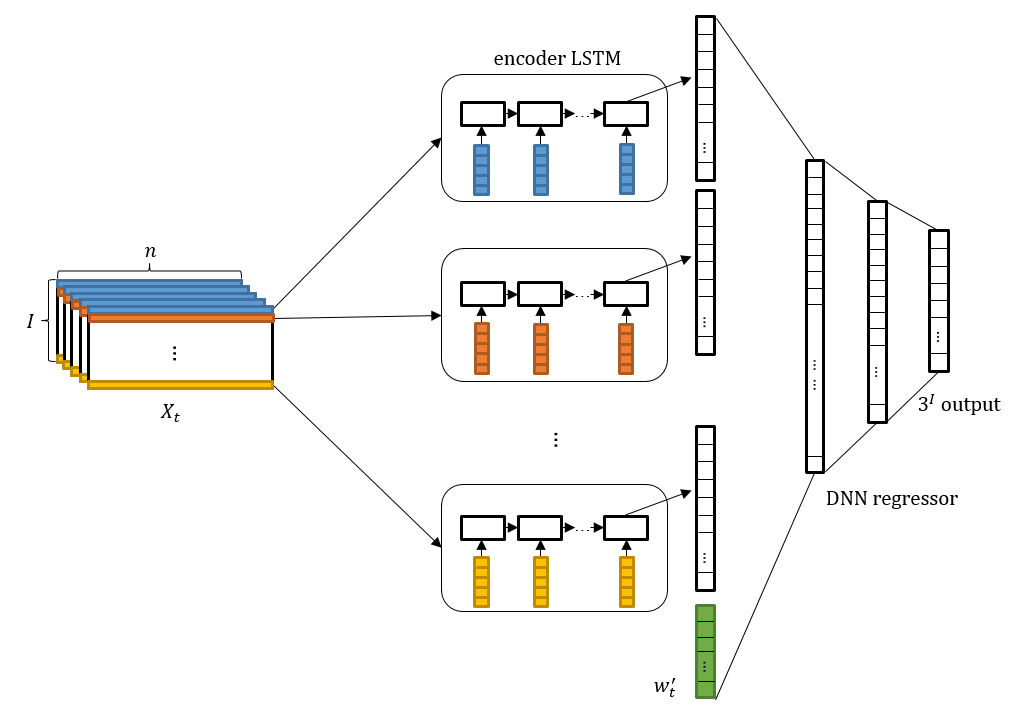}
	\caption{Q-network structure}\label{fig:qnet}
	\end{figure*}

To implement DQN, we need to model the neural network structure for approximating the Q-function of an agent's state and action. We construct a hybrid encoder LSTM-DNN neural network that enables us to approximate the Q-value of an agent's action in our predefined state and action space. First, we train LSTM autoencoder, an unsupervised learning method for compressing sequence data, for identifying latent variables of the historical sequence of predefined technical indicators of assets in the portfolio~\citep{Srivastava:2015}. Then the autoencoder is fitted in the historical pattern of technical indicators of assets, the decoder is removed and the encoder keeps as a standalone model that encodes the technical indicator sequences for assets to lower dimension latent variables. Each asset in the portfolio shares the same encoder LSTM to take the sequence encoding procedure because it is known that a single deep learning model is more effective for learning the feature patterns of different assets than multiple deep learning models that learning individual assets~\citep{Sirignano:2018}. Then, the encoded outputs for each asset are concatenated to create the intermediate output, and this intermediate output is then combined again with the current portfolio weights to use as the input to the DNN. Through this DNN layers, we can obtain the Q-value of each action of the agent. Because these DNN layers extract meaningful features through nonlinear mapping using a multi-layer neural network and conducts a regression for the Q-value, we refer to these layers as the DNN regressor. The overall Q-network structure is as shown in Figure~\ref{fig:qnet}. In summary, the overall DQN algorithm for our approach for deriving the portfolio trading strategy is as follows Algorithm (2). In offline learning, we use this algorithm to build an initial trading strategy fitted on historical data and in online learning, adapt the trading strategy by updating based on the daily observed data in the trading process.

\begin{algorithm}[!htb]
\caption{DQN algorithm for portfolio trading}
\label{alg:DQNPM}
\begin{algorithmic}[1]
\State $F(s)$ : feasible action set in state $s$ 
\State Initialize replay memory $D$
\State Initialize weights of Q-network $\theta$ randomly
\State Initialize weights of target network $\theta' \gets \theta$
\For{episode is sampled by sampling distribution $y \leftarrow g_{\beta}(\cdot)$}
    \State Initialize state $s_0$
    \For{period $t$=0...$T$ in episode $y$}
        \State With probability $\epsilon$ select random $a_t{\in}F(s_t)$ \par
        \hspace{0.3cm} otherwise, $a_t =\begin{cases} \underset{a}{\operatorname*{argmax}} Q(s_t,a;\theta) \hspace{0.2cm}if \hspace{0.2cm}\underset{a}{\operatorname*{argmax}} Q(s_t,a;\theta) \in F(s_t),\\ \textproc{Map}(s_t,\underset{a}{\operatorname*{argmax}} Q(s_t,a;\theta))\hspace{0.3cm}o/w \end{cases}$
        \State Take action $a_t$ and then observe reward $r_t$ and next state $s_{t+1}$
        \State Simulate all actions $a \in F(s_t)$, then observe experience list $L$
        \State Store $L$ in replay memory $D$
        \State Sample random batch of experience list $K$ from $D$
        \State $(s_t,a_t,r_t,s_{t+1})$ is element of experience list in batch, \par
        \hspace{0.3cm} update Q-network from current prediction $Q(s_t,a_t;\theta)$ to target \par \hspace{0.3cm}$z_t =\begin{cases} r_{t}+{\gamma}\underset{a'}{\operatorname*{max}} \hat{Q}(s_{t+1},a';\theta') \hspace{0.2cm}if \hspace{0.2cm}\underset{a'}{\operatorname*{argmax}} \hat{Q}(s_{t+1},a';\theta') \in F(s_{t+1}),\\ r_{t}+{\gamma}\underset{a'}{\operatorname*{max}} \hat{Q}(s_{t+1},\textproc{Map}(s_{t+1},\underset{a'}{\operatorname*{argmax}} \hat{Q}(s_{t+1},a';\theta'));\theta')\hspace{0.2cm}o/w \end{cases}$
        \State Update $\theta$ by minimizing the loss: \par \hspace{0.3cm} $L(\theta)=\frac{1}{\mid K \mid}\sum_{L \in K}\sum_{j \in L}(z_j-Q(s_j,a_j;\theta))^{2}$
    \EndFor
    \State $\theta' \leftarrow \theta$
\EndFor
\end{algorithmic}
\end{algorithm}


\section{Experimental results}\label{sec:Experimental}

In this section, we demonstrate that the DQN strategy (i.e., the trading strategy derived using our proposed DQN algorithm for portfolio trading) can outperform in real-world trading. We conduct a trading simulation for two different portfolio cases using both our DQN strategy and traditional trading strategies as benchmarks, and we verify that the DQN strategy is relatively superior to the other benchmark strategies based on several common performance measures. 

\subsection{Performance measures}\label{sec:performance}

We use three different output performance measures to evaluate trading strategies. The first measure is the cumulative return based on the increase in the portfolio value at the end of the investment horizon relative to the initial portfolio value, as defined as Equation (19): 
\begin{align}\label{eq:model9}
& CR=\frac{P_{t_f}-P_0}{P_0}\times100(\%) \text{, }
\end{align}
where $t_f$ is the final date of the investment horizon and $P_0$ is the initial portfolio value.

The second measure is the Sharpe ratio, as defined in Equation (20): 
\begin{align}\label{eq:model10}
& SR=\frac{E[\rho_t-\rho_f]}{std({\rho_t})}\times\sqrt{252} \text{, }
\end{align}
where $std({\rho}_t)$ is the standard deviation of the daily return rate, $\rho_f$ is the daily risk-free rate (assumed to be 0.01\%), and annualization term (square root of the number of annual trading days) is multiplied. This ratio is a common measure of the risk-adjusted return, and it is used to evaluate not only how high the risk premium is but also how small the variation in the return rate is. 

For the last measure, we use the customized average turnover rate defined as in Equation (21): 
\begin{align}\label{eq:model11}
& AT=\frac{1}{2t_f}\sum_{t=0}^{t_f}\sum_{i=1}^{I}{{\mid}\hat{w}_{t,i}'-w_{t,i}'\mid}\times100 (\%)  \text{. }
\end{align}
The average turnover measures the average rate of change of the portfolio weight vector during the investment horizon. We do not have to consider changes in the cash proportion, so we customize this measure by excluding the change in the weight on cash before and after the agent takes an action. This rate can evaluate the change in the proportions of asset investments. Considering transaction costs, this measure should be low to better apply the trading strategy in the real world. 

\subsection{Data summary}
We experiment with two different three-asset portfolios. The first consists of three exchange traded funds (ETFs) in the US market that track the S\&P500 index, the Russell 1000 index, and the Russell Microcap Index. This type of portfolio was tested in a previous study~\citep{Almahdi:2017}. The second portfolio is a Korean portfolio consisting of the KOSPI 100 index, the KOSPI midcap index, and the KOSPI microcap index. More information for these test portfolios is provided in Table~\ref{tab:portfolios}. 

\begin{table}[!htb]
\begin{adjustwidth}{-1.0in}{-1.0in}
\begin{center}
\begin{threeparttable}
\footnotesize
\caption{Test portfolios}\label{tab:portfolios}
\begin{tabular}{c m{4cm} m{4cm}}
    \hline
    Assets   &  \multicolumn{2}{c}{Portfolio}\\[5pt]
    \cline{2-3}
       & US Portfolio (US-ETF)     & Korean Portfolio (KOR-IDX) \\
    \hline
    Asset 1 & SPDR S\&P 500\tnote{1} & KOSPI 100 index\\
    \hline
    Asset 2 & iShares Russell 1000 Value\tnote{2} & Midcap KOSPI index\\
    \hline
    Asset 3 & iShares Microcap\tnote{3} & Microcap KOSPI index\\
    \hline
\end{tabular}
\begin{tablenotes}
\scriptsize
\item[1] ETF tracks the S\&P500 index
\item[2] ETF tracks the Russell 1000 (mid- and large-cap US stocks) index
\item[3] ETF tracks the Russell microcap index
\end{tablenotes}
\end{threeparttable}
\end{center}
\end{adjustwidth}
\end{table}

We obtain data on the three US ETFs from \textit{Yahoo Finance} and data on the Korean indices from \textit{Investing.com}. Both cases are tested in 2017. The trading strategy for the US portfolio is derived by training on data from 2010 to 2016, and the trading strategy for the Korean portfolio is derived by training on data from 2012 to 2016. 

\subsection{Experiment setting}\label{sec:setting}

Through several rounds of tuning, we derive appropriate hyper-parameters. In particular, the time window size$(n)$ is the most important hyper-parameter, and we adopt the value of 20 among the candidates (5,20,60,120). This time window size and the other tuned hyper-parameters are summarized in Table~\ref{tab:parameter}. 

\begin{table}[h]                           
 \centering
    \begin{tabular}{|c|c!{\vrule width 1pt}c|c|}
    \hline
     \rowcolor{lightgray} \textbf{hyper-parameter}   &   \textbf{value}    &   \textbf{hyper-parameter}   &   \textbf{value}    \\ \hline
     time window size $(n)$    &   20  &  replay memory size &   2000      \\ \hline          
     learning rate $(\alpha)$   &   1e-7 & number of epochs    &   500\\ \hline
     distribution parameter $(\beta)$ &   0.3  & discount factor $(\gamma)$  &   0.9   \\ \hline
     DNN input dimension  &   64  &    batch size    &   32       \\ \hline
     DNN layer       &  2  &    encoder LSTM layer    &   1 \\ \hline
     DNN 1st layer dimension       &   64 &   hidden dim of encoder LSTM    &   128 \\ \hline
     DNN 2nd layer dimension       &   32 &   output dim of encoder LSTM    &   20 \\ \hline
    \end{tabular}
    \caption{hyper-parameter summary}
    \label{tab:parameter}                            
\end{table}

In the experiment, we also need to set trading parameters, such as the initial portfolio value and the trading size. We set the initial portfolio value as one million in both portfolio cases (e.g., 1M $USD$ for the US portfolio and 1M $KRW$ for the Korean portfolio). Similarly, we set the trading size as ten thousand in both portfolio cases (e.g., 10K $USD$ trading size for the US portfolio case and 10K $KRW$ for the Korean portfolio case). We set the transaction cost rate for buying and selling in both the US and Korean markets as 0.25\%. In both cases, the initial portfolio is set up as an equally weighted portfolio, in which every asset and cash has the same proportion.

\subsection{Benchmark strategy}\label{sec:benchmark}

To evaluate our DQN strategy, we compare it to some traditional portfolio trading strategies. The first strategy is a buy-and-hold strategy $(B\&H)$ that does not take any action but rather holds the initial portfolio until the end of the investment horizon. The second strategy is a randomly selected strategy $(RN)$ that takes action within the feasible action space randomly in each state. The third strategy is a momentum strategy $(MO)$. This strategy buys assets whose values increased in the previous period and sells assets whose values decreased in the previous period. However, if it cannot buy all assets with increased values, it gives buying priority to assets whose values increased more. If it is unable to sell assets whose values decreased, it simply holds the assets. The last strategy is a reversion strategy $(RV)$, which is the opposite of the momentum strategy. This strategy sells assets whose values increased in the previous period and buys assets whose values decreased in the previous period. However, if it cannot buy all of the assets whose values decreased, it gives buying priority to the assets whose values decreased more. If it is unable to sell the assets whose values increased, then it simply holds the assets. 
 
\subsection{Result}\label{sec:result}

\begin{figure}[!htb]%
    \centering
    \subfloat[]{{\includegraphics[width=0.8\textwidth]{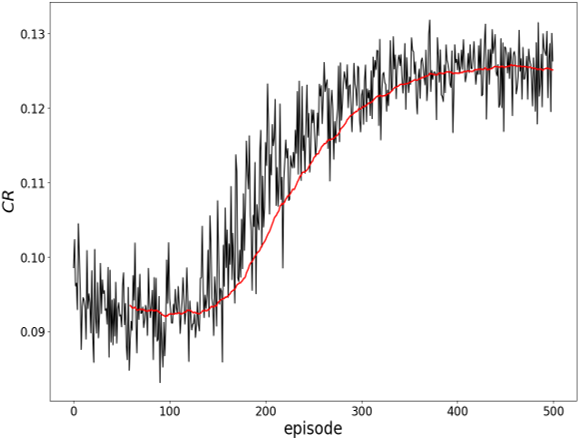}}}%
    \qquad
    \subfloat[]{{\includegraphics[width=0.8\textwidth]{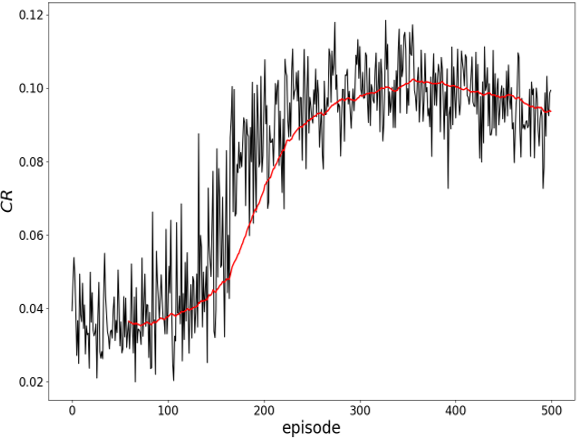}}}%
    \caption{Cumulative return rate as the learning episode continues (a) US portfolio, (b) Korean portfolio}\label{fig:return}%
\end{figure}

We derive a trading strategy for both portfolio cases using DQN. For both cases, we identify the increase in the cumulative return over the investment horizon of the test period as episode learning continues. Figure~\ref{fig:return} shows the trend in the cumulative return performance over the learning episodes in both cases.

Table~\ref{tab:vs_map} shows the number of changes in the phase of trading direction (i.e., the number of changes from selling to buying or vice versa, except for holding) before and after applying the mapping function. A trading strategy that has frequent changes in the phase of trading action implies an unreasonable trading strategy because these changes deteriorate trading performance by incurring meaningless transaction costs. Through the experimental result, we identify that when the mapping function is applied to the DQN strategy, the number of changes in the phase of the trading direction decreases relative to when it is not applied. Following by decreasing the number of changes in the phase of the trading direction, the cumulative return of the trading strategy increases by 6.68\% relative to the one without applied the mapping function in the US portfolio case. Likewise, in the Korean portfolio case, the cumulative return of the trading strategy increases by 10.83\% relative to the one without applied the mapping function. Therefore, we demonstrate that the mapping function contributes to deriving a reasonable trading strategy.

\begin{table}[!htb]                           
 \centering
\begin{tabular}{c|ccc|ccc}
  \toprule
  \rowcolor{gray!20}  & \multicolumn{3}{c}{US portfolio} \vline & \multicolumn{3}{c}{Korean portfolio} \\ \cmidrule{2-7}
  \rowcolor{gray!20} \multirow{-2}{*}{mapping function} & asset 1 & asset 2 & asset 3 & asset 1 & asset 2 & asset 3\\
   \midrule 
   without & 83& 107 & 64 & 74 & 128 & 63 \\
   with & 54 & 40 & 37 & 42 & 68 & 40 \\
   \hline
   difference & -34.9\% &-62.6\%&-42.1\%&-43.2\%&-46.8\%&-36.5\%\\
 \bottomrule
\end{tabular}
\caption{The number of changes in the phase of trading direction for DQN strategy without/with the mapping function in the two test portfolio cases}
\label{tab:vs_map}
\end{table}

\begin{figure}[!hb]%
    \centering
    \subfloat[]{{\includegraphics[width=0.9\textwidth]{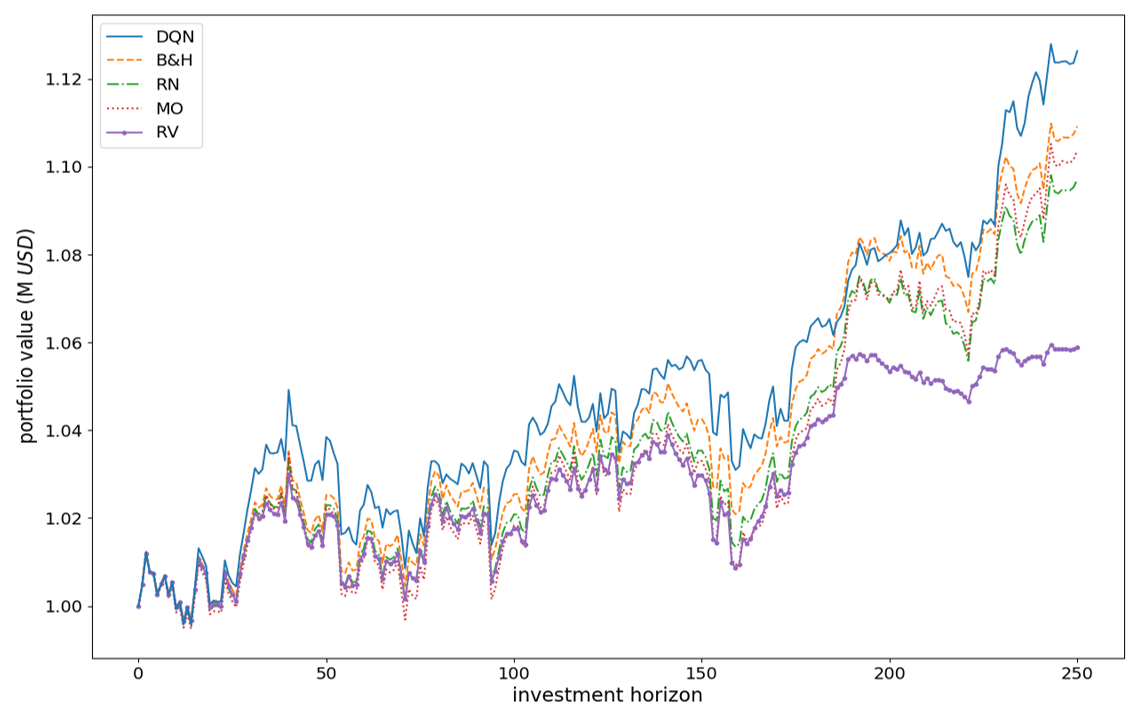}}}%
    \qquad
    \subfloat[]{{\includegraphics[width=0.9\textwidth]{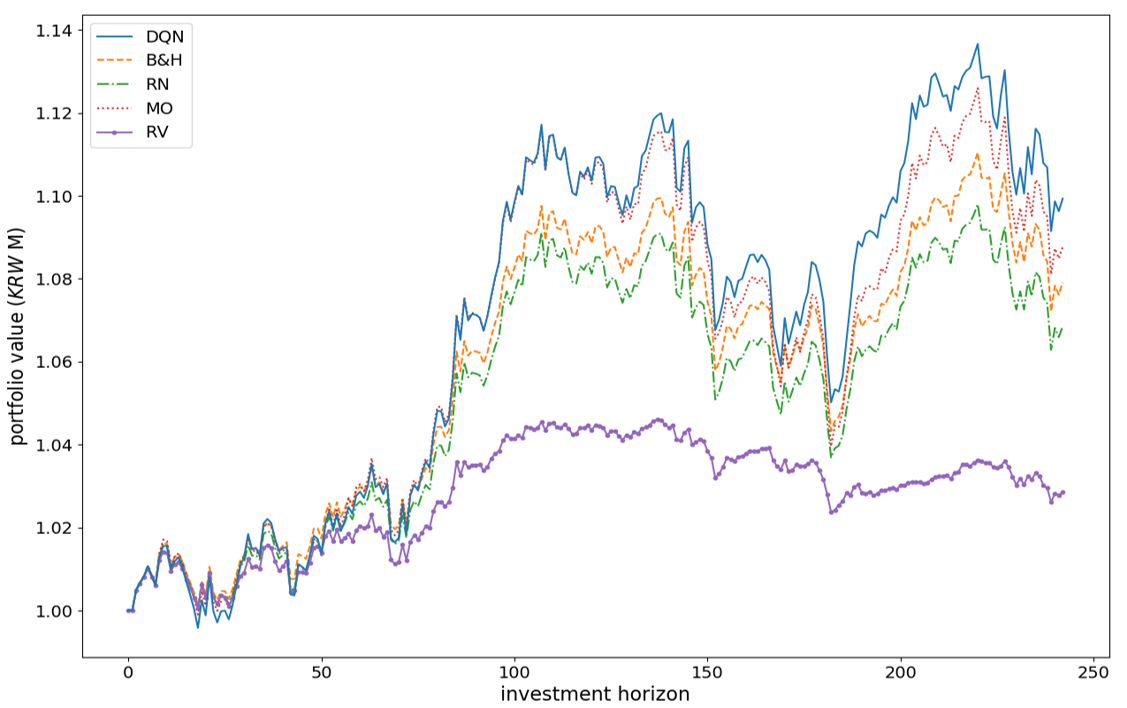}}}%
    \caption{Comparative portfolio value results for the DQN and benchmark strategies for (a) US portfolio, (b) Korean portfolio}\label{fig:trend}%
\end{figure}

Figure~\ref{fig:trend} shows the portfolio value trend when applying the DQN strategy and the benchmark strategies in the US and Korean portfolio cases. In the US portfolio case, we observe that the DQN strategy outperforms the benchmark strategies for most of the test period. The final portfolio value of the DQN strategy is 15.69\% higher than that of the B\&H strategy, 33.74\% higher than that of the RN strategy, 21.81\% higher than that of the MO strategy, and 114.47\% higher than that of the RV strategy. Likewise, in the Korean portfolio case, we observe that the DQN strategy outperforms the benchmark strategies for most of the test period. The final portfolio value of the DQN strategy is 25.52\% higher than that of the B\&H strategy, 34.99\% higher than that of the RN strategy, 13.22\% higher than that of the MO strategy, and 247.91\% higher than that of the RV strategy.

Table~\ref{tab:perform} summarizes the output performance measure results when using DQN and the benchmark strategies in both portfolio cases. This table shows that the DQN strategy has the best cumulative return and Sharpe ratio performances for the US portfolio, and this strategy has the lowest turnover rate except for the B\&H strategy, which has no turnover rate. In the Korean portfolio case, the DQN strategy also has the best cumulative return and Sharpe ratio performances. Moreover, the DQN strategy has the lowest turnover rate except for the B\&H strategy. Given that the B\&H strategy does not incur any transaction costs during the investment horizon, it is a remarkable achievement that the DQN strategy outperforms the B\&H strategy in terms of the cumulative return and Sharpe ratio.

\begin{table}[!htb]                           
 \centering
\begin{tabular}{c|ccc|ccc}
  \toprule
  \rowcolor{gray!20}  & \multicolumn{3}{c}{US portfolio} \vline & \multicolumn{3}{c}{Korean portfolio} \\ \cmidrule{2-7}
  \rowcolor{gray!20} \multirow{-2}{*}{strategy} & $CR$ & $SR$ & $AT$ & $CR$ & $SR$ & $AT$\\
   \midrule 
   \small{$B\&H$} & 10.921\%& 1.308 & \textbf{0.000\%} & 7.913\% & 0.890 & \textbf{0.000\%} \\
   $RN$* & 9.446\% & 1.143 & 1.029\% & 7.358\% & 0.381 & 1.032\% \\
   $MO$ & 10.372\% & 1.114 & 1.368\% & 8.773\% & 0.840 & 1.233\% \\
   $RV$ & 5.891\% & 0.642 & 1.404\% & 2.855\% & 0.147  & 1.370\% \\
   $\textbf{DQN}$ & \textbf{12.634\%} & \textbf{1.382} & \textbf{0.954\%} & \textbf{9.933\%} & \textbf{0.946} & \textbf{0.989\%} \\
 \bottomrule
\end{tabular}
\begin{tablenotes}
\scriptsize
\item[1] *performance of $RN$ is average from 30 samples 
\end{tablenotes}
\caption{Output performance measure values for our DQN strategy and the benchmark strategies in the two test portfolio cases}
\label{tab:perform}
\end{table}

\section{Conclusion}\label{sec:Conc}

The main contribution of our study is applying the DQN algorithm to derive a portfolio trading strategy on the practical action space. However, applying DQN to portfolio trading has some challenges. To overcome these challenges, we devise a DQL model for trading and several techniques. First, we introduce a mapping function for handling infeasible actions to derive a reasonable trading strategy. Trading strategies derived from RL agents can be unreasonable to apply in the real world. Thus, we apply a domain knowledge rule to develop a trading strategy with an infeasible action mapping constraint. As a result, this function works well, and we can derive a reasonable trading strategy. Second, we design DQL agent and Q-network for considering multi-asset features and derive a multi-asset trading strategy in the practical action space, determining the trading direction of the asset, by overcoming the dimensionality problem. Third, we relax the data shortage issue for deriving well-fitted multi-asset trading strategies by introducing a technique that simulates all feasible actions and then updating the trading strategy based on the experiences of these simulated actions. 

The experimental results show that our proposed DQN strategy is a superior trading strategy relative to benchmark strategies. Based on the results of the cumulative return and the Sharpe ratio, the DQN strategy is more profitable with lower risk than other benchmark strategies. In addition, based on the results of the average turnover rate, the DQN strategy is more suitable for application in real-world trading than benchmark strategies.

However, our proposed methodology still has a limit of scalability, arising the dimensionality problem if the number of assets in the portfolio very large. Furthermore, in our study, the reward of the MDP model is optimized only for returns and not for risk. Nevertheless, the contributions of this study are still valuable because of the novel techniques for expressing the practical applicability of the portfolio trading strategy. By responding to the limit of the current study, we will try to devise the method for resolving these limitations in future research.

\end{document}